\documentstyle[12pt]{article}

\begin{document}

\LaTeX{}

\begin{center}
The plasma-solid transition\bigskip

V.Celebonovic$^{1}$ and W.D\"appen$^{2}$

$^{1}$Institute of Physics, Pregrevica 118,11080 Zemun-Beograd, Yugoslavia

$^{2}$Dept. of Physics and Astronomy, USC, Los Angeles, 
CA 90089-1342, USA\bigskip
\end{center}

Abstract: Using a criterion proposed by Salpeter and standard solid-state
physics, we have determined conditions for the occurrence of the
plasma-solid transition. Possible astrophysical applications are discussed.

\medskip

\begin{center}
Introduction
\end{center}

Phase transitions of various kinds have been, and still are, actively
contributing to important changes occurring in 
the world around us. The aim of the present paper
is a contribution to the study of the conditions for the occurrence of
the so-called ``plasma-solid (PS)'' transition.

Simple physical reasoning shows that the PS~transition must
certainly occur in
various astrophysical settings, such as in 
proto-planetary and proto-stellar
clouds. On the one hand, it is widely
known that the universe contains different kinds of plasmas. 
On the other hand, it is
equally widely known 
that solid objects also exist in the universe. This implies that
there must exist a physical region where a transition between
the two regimes takes place.

In our preliminary work on the PS~transition $\left[ 1\right]$, we have
determined the conditions for its occurrence in two simple idealized systems:
a pure Fermi-Dirac and a pure Bose-Einstein gas. The object of the
present calculation is again a determination of the conditions for the
occurrence of the PS~transition, but this time 
on a more realistic footing. Our
starting point is the criterion for the occurrence of the PS~transition
(although it was not called by that name) proposed by Salpeter $\left[
2\right]$.

Salpeter's paper was devoted to a thorough discussion of a zero-tempera-
ture plasma. In its third part, he considered 
a system of positive ions of given
charge and mass, rigidly fixed in the nodes of a perfect crystal
lattice,
and went on to {\it estimate} the zero point energy of the ions.
He showed
that the behavior of such a system can be described by the ratio 

\begin{equation}
\label{(1)}\ f=\frac{E_{z,p}}{E_{C}} \ ,
\end{equation}
where $E_{z,p}$ denotes the zero point energy of the ions and $E_{C}$ is the
Coulomb energy. According to the analysis presented in $\left[ 2\right] $, a
PS~transition occurs for $f=1$. Calculations reported in $\left[ 2\right] $
were estimates, and strictly valid only for $T=0$~K. Our calculations
reported in the following section go beyond those assumptions in at
least two aspects: (i) we have not used estimates but exact
calculations, and (ii) our results take into account the influence of
temperature.
\medskip

\begin{center}
Calculations
\end{center}

Finding a general expression for the energy of a real solid that takes
into account most (if not all) of its characteristics is a formidable
problem in solid state physics (for example $\left[ 3\right] ,\left[
4\right] $ ). Various approximations to the complete solution exist; one
of them is the Debye model of a solid. Within the Debye model,
the energy per mole
of a solid is given by the following expression $\left[ 3\right] $:

\begin{equation}
\label{(2)}\ E=N\left\{ u\left( v\right) +9nk_{B}T\left[ \left( \frac{T}{\theta }\right)
^{3}\int_{0}^{\theta /T}\left( \frac{1}{2}+\frac{1}{\exp \left( \xi \right)
-1}\right) \xi ^{3}d\xi \right] \right\} \ ,
\end{equation}
where
$N$ denotes the number of elementary cells in a mole of the material,
$n$ the number of particles per elementary cell,
$u(v)=\frac{U}{N}=$ the static lattice energy per cell,
$\theta$ the Debye temperature,  
$k_{B}$ the Boltzmann constant, and in the following we will use 
the convention $k_{B}=1$. 
The second integral in Eq.~(2) can be solved as $\left[ 5\right] $

\begin{equation}
\label{(3)}\ I_{1}=\int_{0}^{x}\frac{t^{n}dt}{e^{t}-1}=x^{n}\left[ \frac{1}{n}-\frac{x}{%
2\left( n+1\right) }+\sum_{k=1}^{\infty }\frac{B_{2k}x^{2k}}{\left(
2k+n\right) \left( 2k\right) !}\right] 
\end{equation}

under the condition 
$ n\succeq 1,\left| x\right| \prec 2\pi $. 
Here, the symbol $B_{2k}$ denotes Bernoulli's 
numbers. It follows from Eq.~(2) that the energy
per particle of a solid within the Debye model is given by
\begin{equation}
\label{(4)}\ E_{p}=9T\left( \frac{T}{\theta }\right) ^{3}I 
\end{equation}
where 
\begin{equation}
\label{(5)}\ I=\frac{1}{2}\int_{0}^{\theta /T}\xi ^{3}d\xi +\int_{0}^{\theta /T}\frac{\xi
^{3}d\xi }{e^{\xi }-1} 
\end{equation}
Using Eq.~(3) in Eq.(5), one finally gets

\begin{equation}
\label{(6)}\ \left( \frac{T}{\theta }\right) ^{3}I=\frac{1}{3}+\sum_{k=1}^{\infty }\frac{%
B_{2k}}{\left( 2k+3\right) \left( 2k\right) !}\left( \frac{\theta }{T}%
\right) ^{2k} \ ,
\end{equation}
which implies that

\begin{equation}
\label{(7)}\ E_{p}=3T\left[ 1+3\sum_{k=1}^{\infty }\frac{B_{2k}}{(2k+3)\left( 2k\right) !}%
\left( \frac{\theta }{T}\right) ^{2k}\right]  
\end{equation}
or, expanding explicitly

\begin{equation}
\label{(8)}\ E_{p}\cong 3T\left[ 1+\frac{1}{20}\left( \frac{\theta }{T}\right) ^{2}-\frac{%
1}{1680}\left( \frac{\theta }{T}\right) ^{4}+..\right] \ .
\end{equation}
The energy of a Fermi-Dirac gas of particles of mass $m$ and spin $s$
enclosed in a volume $V$ is given by $\left[ 6\right] $

\begin{equation}
\label{(9)}\ E_{e}=\frac{gVm^{3/2}}{2^{1/2}\pi ^{2}\hbar ^{3}}F_{3/2}\left( \beta \mu
\right) 
\end{equation}
All the symbols in this equation have their usual meaning, in
particular, $g=2s+1$, and $F_{3/2}\left( \beta \mu \right) $
is a particular case of a Fermi integral

\begin{equation}
\label{(10)}\ F_{k}\left( \beta \mu \right) =\int_{0}^{\infty }\frac{\epsilon
^{k}d\epsilon }{1+\exp \left[ \beta \left( \epsilon _{{}}-\mu \right)
\right] } \ .
\end{equation}
Eq.~(10) can be developed as $\left[ 7\right] $

\begin{eqnarray}
\ F_{k}\left( \beta \mu \right)  &=&\frac{\mu ^{k+1}}{k+1}+  \label{(11)} \\
&&T\sum_{n=0}^{\infty }\frac{1}{n!}\frac{\partial ^{n}}{\partial \mu ^{n}}%
\left( \mu ^{k}\right) \left[ 1-\left( -1\right) ^{n}\right] \left(
1-2^{-n}\right) T^{n}\Gamma \left[ n+1\right] \varsigma \left[
n+1\right] \ ,
\nonumber
\end{eqnarray}
where $\Gamma $ and $\varsigma $ are the gamma function and Riemann zeta
function, respectively.

Expanding to fourth order and inserting the temperature dependence of the
chemical potential, one gets the following result for the energy per particle
of the electron gas

\begin{equation}
\label{(12)}\ \frac{E_{e}}{N}=\frac{gm^{3/2}}{\pi ^{2}\hbar ^{3}}
\frac{2^{1/2}}{5}\mu
_{0}^{5/2}\left[ 1+\frac{5\pi ^{2}}{12}\left( \frac{T}{\mu _{0}}\right) ^{2}-%
\frac{\pi ^{4}}{36}\left( \frac{T}{\mu _{0}}\right) ^{4}\right]  
\end{equation}
The chemical potential of the electron gas at $T=0$~K is denoted here
by $\mu _{0}=An^{2/3}$,
with $A=\left( 3\pi ^{2}\right) ^{2/3}{\hbar ^{2}}/{2m}$.

Introducing 
$B=({2^{1/2}}/{5}){gm^{3/2}}/({\pi ^{2}\hbar ^{3}})$
simplifies the notation in Eq.~(12). 

\begin{center}
\bigskip Discussion
\end{center}

Salpeter's ratio defined in Eq.~(1) can be formed using Eqs.~(8)
and~(12).
The result obtained this way can be expressed as follows

\begin{equation}
\label{(13)}\ f\simeq \frac{9}{140}\frac{\mu _{0}^{3/2}}{BT^{3}}\frac{1680T^{4}+84\left(
T\theta \right) ^{2}-\theta ^{4}}{36\mu _{0}^{4}+15\left( \pi \mu
_{0}T\right) ^{2}-\left( \pi T\right) ^{4}}  
\end{equation}
Imposing the condition $f=1$ on Eq.(13),
it becomes possible to determine the
values of the parameters of a PS~transition. Solving Eq.~(13) for the Debye
temperature $\theta $, and introducing in the solution the density
dependence of the chemical potential at $T=0$~K one obtains the following
expression
\begin{eqnarray}
\theta &=&\left( \frac{2}{3}\right) ^{1/2}[63T^{2}+\frac{1}{A^{2}n^{4/3}}(%
\sqrt{7An^{2/3}}(1107n^{2}A^{3}T^{4}+  \label{(14)} \\
&&5\sqrt{An^{2/3}}(ABn^{2/3}\pi ^{4}T^{7}-36A^{5}Bn^{10/3}T^{3}-15A^{3}B\pi
^{3}n^{2}T^{5}))^{1/3}]^{1/2} \ . \nonumber
\end{eqnarray}
This equation links the relevant
parameters of a plasma (number density and
temperature) with those of the solid (Debye temperature) that can condense 
from it. We have thus
obtained the equation of state of a system undergoing a PS~transition.
Apart
from being interesting from the point of view of pure statistical
physics, this result can find astrophysical applications in studies of
dust and gas clouds. For recent examples of observational
studies of such clouds, 
see for example $\left[ 8\right] $ $\left[ 9\right]$.

The number density and the temperature of an astrophysical 
cloud can be determined from
observations using the methods of plasma physics, such as the analysis
of spectral line broadening, or in some cases, {\it in-situ} 
measurements of 
interplanetary gas and dust. Once these values are known for a given
cloud, the Debye temperature of the solid material that 
can condense from it is
given by Eq.~(14). The Debye temperature is a unique characteristics 
of every
solid, which gives hope that the method discussed here could lead to a
possibility to determine the chemical composition of a condensing
proto-planetary system. Such a determination would of course be 
only indirect, because
after calculating the value of $\theta $ from Eq.~(14), one would have to
rely on an identification of 
the material (or a mixture of materials) based on the corresponding
calculated value.

Consider Eq.(14) in the limiting case 
$T\rightarrow0$,\ $\mu_0\rightarrow0$.
Astronomically, 
this limit corresponds to an interstellar cloud sufficiently far 
away 
from any bright star. Developing Eq.~(14) 
into series in the chemical potential 
and retaining just the first two terms, one gets the following expression for 
the Debye temperature
\begin{equation}
\label{(15)}\ \theta  = K_1 T^{7/6} \mu_{0}^{-1/2}+K_2 T^{5/6} \mu_{0}^{1/2}
\end{equation}
The numerical values of the constants in Eq.~(15) can be determined by
developing Eq.~(14) into a series. By doing this, it follows that
\newline  
$\ K_1$=$(2/3)^{1/2} 5^{1/6} 7^{1/4} \pi^{2/3} B^{1/6}$ and 
$\ K_2$=$3^{3/2} 7^{3/4} 2^{-1/2} \pi^{-2/3} (5 B)^{-1/6} $. \newline
Here, the constant $B$ is the one 
defined after Eq.~(12). For sufficiently small values of 
the chemical potential, Eq.~(15) can be simplified into
\begin {equation}
\label {(16)}\ \theta = K_1 T^{7/6} \mu_{0}^{-1/2}
\end {equation}
Inserting the values of all the constants which occur in Eq.~(16),
one finally 
arrives at
\begin{equation}
\label{(17)}\ \theta = 50.8 T^{7/6} n^{-1/3}
\end {equation} 
This expression has two possible applications.
It can be used for the 
calculation of $\theta$ if the temperature and the number density of the cloud 
are known. Typical values of the temperature and density of the molecular 
clouds in our Galaxy are $20<T[{\rm K}]<60$ and $10^3<n({\rm cm}^{-3}) <10^5$.
Inserting in Eq.~(17) 
$n=2000 {\rm cm}^{-3}$ and $T=43$~K gives $\theta = 320$~K,
which is nearly the 
experimentally known value of the Debye temperature of the chemical element 
magnesium. 
This result could be interpreted as indication that in a molecular
cloud, grains of Mg can condense under these densities and 
temperatures.
In a similar way, the pair of values 
$ n = 10^4 {\rm cm}^{-3}$, $T=120$~K would lead to
$ \theta = 628$~K. Such a 
value is only slightly higher than the experimental 
value for the element Si. 
Note that this last result has a direct significance for the interpretation 
of observations, because emission from silicate grains has indeed
been detected [10].
Turning the argument around, 
if solid particles are observed in a cloud,
and if their chemical composition can be determined,
it becomes possible to calculate 
the Debye temperature from the principles of solid state physics.
If, in addition, temperature 
is known from spectroscopy, one can determine the value of the 
chemical potential, and finally, the number density of a cloud from 
Eq.~(16).
Further work along these lines is in progress and 
will be discussed elsewhere.\\
{\it Acknowledgment}: One of us (W.D.) was
supported in part by the grant AST-9987391 of the
National Science Foundation (USA).
\newpage
\begin{center}
References
\end{center}
$\left[ 1\right] $ V. Celebonovic and W. D\"appen: 
in:Contributed papers 
of the 20$^{th}$ SPIG Conference, (ed. Z.Lj.Petrovic, M.M.Kuraica, 
N.Bibic 
and G.Malovic),p.527, 
published by the Inst. of Physics, Faculty 
of Physics and INN. Vinca,\newline Beograd,Yugoslavia (2000) 
(preprint available at LANL: astro-ph/0007337).

$\left[ 2\right] $ E.E.Salpeter: Astrophys.J., {\bf 134},669 (1961) .

$\left[ 3\right] $ M.Born and K.Huang: Dynamical theory of crystal
lattices,\newline OUP, Oxford (1968) .

$\left[ 4\right] $ A.Davydov: Theorie du Solide, Editions ''Mir'', Moscou
(1980) .

$\left[ 5\right] $ M.Abramowitz and I.A.Stegun: Handbook of mathematical
functions, Dover Publications Inc., New York (1972).

$\left[ 6\right] $L.D.Landau and E.M.Lifchitz: Statistical
Physics,Vol.1, Nauka,\newline Moscow (1976) (in Russian).

$\left[ 7\right] $ V.Celebonovic: Publ.Astron.Obs.Belgrade,{\bf 
60},16,(1998).\newline(preprint available at LANL: astro-ph/9802279).

$\left[ 8\right] $ N.E.Jessop and D.Ward-Thompson: 
Mon.Not.R.astr.Soc. (in press) (preprint available at LANL: astro-ph/0012095). 

$\left[ 9\right] $ E.Grun,H.Krueger and M.Landgraf: in:The 
Heliosphere at Solar Minimum: The Ulysses Perspective, eds. 
A.Balogh, R.Marsden, E.Smith,Springer 
Praxis,Heidelberg (in press) 
(preprint available at LANL: astro-ph/0012226). 

$\left [ 10\right] $ D.Cesarsky,A.P.Jones,L.Lequeux,L.Verstrate:
Astron.Astrophys.,
\newline{\bf 358},708 (2000).

\end{document}